\documentclass[aps,prl,twocolumn,superscriptaddress,groupedaddress,a4paper]{revtex4-1}
\usepackage{graphicx}
\usepackage{geometry}
\usepackage{color}

\begin{document}

%\preprint{}

\title{Comment on ``Large Fluctuations for Spatial Diffusion of Cold Atoms''} % \\ \vspace{1cm}

\author{Igor Goychuk}
% \email{igoychuk@uni-potsdam.de, corresponding author}
 
\affiliation{Institute of Physics and Astronomy, University of Potsdam, 
Karl-Liebknecht-Str. 24/25,
14476 Potsdam-Golm, Germany}

\begin{abstract}

A comment on the Letter by E. Aghion, D. Kessler, and E. Barkai,
Phys. Rev. Lett. \textbf{118}, 260601 (2017). 
An important criterion on finite 
kinetic temperature of the system of cold atoms is established. 
It is shown that the kinetic temperature becomes infinite in Fig. 1 of the commented paper
in the course of time, i.e. the considered model system becomes
asymptotically infinitely hot.  Moreover, within this model the 
behavior of the squared width of the spatial distribution of atoms at the half of its maximum is very
different from the variance of the particle positions. In particular, in the discussed Fig. 1
the former one increases sub-ballistically in time, while the variance grows super-ballistically, which corresponds to a heating phase. This leads to a profound ambiguity in definition and classification of anomalous diffusion. All in all, the model in the 
commented paper simply does not fit to experiments with cold atoms.

\end{abstract}

\maketitle

In a recent Letter \cite{Non}, the authors claim to study a system of cold atoms in a non-heating
phase. Below, I show that kinetic temperature of these ``cold'' atoms becomes infinite in 
 Fig. 1 of \cite{Non} in the course of time.

Indeed, the stochastic dynamics in their Eq. (1) is always far away from thermal equilibrium. This is so because the noise and frictional terms in their Eq. (1) are not related by the fluctuation-dissipation theorem \cite{Kubo}. Hence, the equilibrium velocity distribution \textit{never} exists in such a system. What they incorrectly name equilibrium distribution is the steady state solution of  the Fokker-Planck equation
\cite{Kubo,Pitaevski}
\begin{eqnarray}
\frac{\partial P(v,t)} {\partial t}=
\frac{\partial } {\partial v}\left [  D e^{-U(v)/D}
\frac{\partial } {\partial v} e^{ U(v)/D} P(v,t) \right ],
\end{eqnarray}
which in \cite{Kessler10,Non} corresponds to a fictitious ``velocity potential''
$U(v)=(1/2)\ln(1+v^2)$.
This steady state solution reads obviously \cite{Kessler10},
\begin{eqnarray}\label{st}
P_{\rm st}(v)=C e^{-U(v)/D}=\frac{C}{(1+v^2)^{1/(2D)}},
\end{eqnarray}
where $C=\Gamma(1/(2D))/[\sqrt{\pi}\Gamma(1/(2D)-1/2)]$ is the normalization constant 
for $D<1$, and for $D\geq 1$ the steady-state density is not normalizable.
%As the authors correctly mention below their Eq. (3) (with a small inaccuracy),
%the steady state function $P_{\rm st}(v)$ is not normalizable for $D\geq 1$ (they write $D>1$), %and hence then the steady (which they name equilibrium) state does not exist. 
The authors name the parameter regime $D>1$ the heating phase and 
do not consider it further.
However, what is the steady-state kinetic temperature $T_{\rm kin}$ 
of the particles described
by their equation (1)? In accordance with the basic principles of statistical physics it 
can be defined by the mean kinetic energy of the particles provided that 
$\langle v(t)\rangle=0$ (the case) as 
$k_BT_{\rm kin}(t)=M \langle v^2(t)\rangle $ ($M=1$, $k_B=1$ in their paper), where 
$\langle v^n(t)\rangle=\int_{-\infty}^{\infty} v^n P(v,t)dv$, in the limit $t\to\infty$.
This is a standard definition of the kinetic temperature extended beyond equilibrium 
\cite{Kubo,Pitaevski,SieglePRL}.
From this and Eq. (2) it immediately follows that both the \textit{steady-state} mean kinetic
energy and the corresponding kinetic temperature are \textit{\bf infinite} for $D\geq 1/3$.
Indeed, 
\begin{eqnarray}
T_{\rm kin}(\infty)=\langle v^2(\infty)\rangle =\frac{D}{1-3D}
\end{eqnarray}
for $D<1/3$, and $T_{\rm kin}(\infty)=\langle v^2(\infty)\rangle =\infty$ otherwise. Apart from 
the temperature interpretation, this is the same expression as Eq. (4) in \cite{Kessler10}.
Unfortunately, anything is stated in \cite{Non} on that for any asymptotically finite kinetic temperature
one must fulfill this very important, crucial condition $D<1/3$. 
For example,  in the experimental work \cite{Douglas} and in Fig. 2 of the  
minireview \cite{Lutz}, $D=(q-1)/2\approx 0.19\div 0.198$, 
with $q\approx 1.38\div 1.396$ therein, i.e. it obeys this condition.
In fact, in Ref. \cite{Kessler10}
the authors show in Eq. (15) that $\langle v^2(t)\rangle $, 
and hence also kinetic temperature, grows algebraically in
time for $1/3<D<1$,
\begin{eqnarray}
T_{\rm kin}(t)=\langle v^2(t)\rangle \propto t^{3/2-1/(2D)}\;.
\end{eqnarray}
Hence, in Fig. 1 of \cite{Non} for
$D=0.4>1/3$, the particles heat up to the infinity. Therefore,
they cannot be considered cold, 
contrary to what is stated in \cite{Non}, even in its title. As a matter of fact, the parameter regime
of continuous heating starts from $D\geq 1/3$, and not from $D>1$, as misleadingly stated in \cite{Non}.
Important to note that for $D\geq 1/3$ a popular operational 
definition of the effective temperature $T_{\rm eff}$ by relating it to the width
of $P(v)$ at its half-maximum
loses any sense within the model of Refs. \cite{Non,Kessler10} and Eq. (\ref{st})
because in this parameter regime 
it spectacularly contradicts to a commonly accepted,
textbook  meaning of the kinetic
temperature.
%  and became then a complete nonsense. 
The validity of this comment can be easily seen from the result on the spatial variance growth
in Eqs. (6), (7) of \cite{Non} yielding 
\begin{eqnarray}\label{diff}
\langle \delta x^2(t)\rangle \propto t^{\alpha(D)}
\end{eqnarray}
with $\alpha(D)=7/2-1/(2D)$
for $1/5< D<1$. Hence, super-diffusion is sub-ballistic for $1/5<D<1/3$ with $\alpha(D)$ gradually
growing from $\alpha=1$ at $D\leq 1/5$ to $\alpha=2$ at $D=1/3$. For $D>1/3$, 
when the particles are heated up continuously, it becomes super-ballistic. Obviously, in this
heating regime 
\begin{eqnarray}
\langle \delta x^2(t)\rangle \propto T_{\rm kin}(t) t^2,
\end{eqnarray}
which corresponds to ballistic diffusion with algebraically growing temperature. Earlier,
similar hyperdiffusive result was found in Ref. \cite{SieglePRL} within a very different model,
where the kinetic temperature increases only transiently. Furthermore, even if the
regime $D>1$ was not studied in Ref. \cite{Non}, Eq. (15) of \cite{Kessler10}
implies that in this case one obtains %universally (within this model for $D\geq 1$)
the Richardson type diffusion
\begin{eqnarray}
\langle \delta x^2(t)\rangle \propto t^3\;.
\end{eqnarray}

\begin{figure}
\resizebox{0.9\columnwidth}{!}{\includegraphics{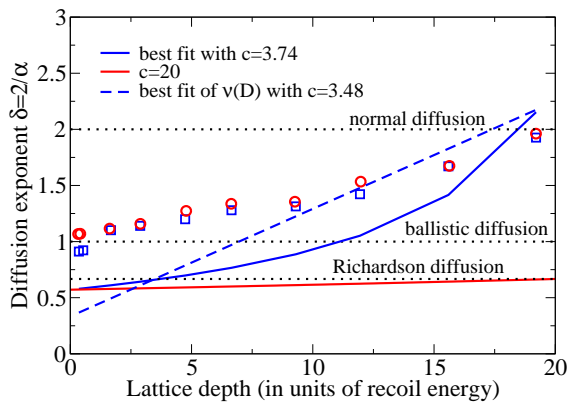}}
\caption{Dependence of the scaling exponent $\delta$ on the optical lattice depth $U_0/E_R$
in units of recoil energy.
Blue squares and red circles are the data extracted from Fig. 3 of Ref. \cite{Sagi12} using
 Engauge Digitizer 10.1. Blue squares correspond to the scaling exponent $\delta$ (which is denoted 
 $\alpha$ in \cite{Sagi12}) extracted from the width of the probability distribution, while
 the red circles correspond to the data fit from a measure for self-similarity used in \cite{Sagi12}). An excellent agreement
 between these two experimental measures in \cite{Sagi12}, except for two outliers for small $U_0$, 
 confirms that
 the probability density is self-similar. The fit with the model in \cite{Non} yields the best fit
 value $c\approx 3.74$ in $D=c E_R/U_0$ (full blue line using $\delta(D)$), or 
 $c\approx 3.48$ (broken blue line using $\nu(D)$). Even the best fit is clearly unacceptable. 
 Notice that below the line $\delta=2/3$ of Richardson diffusion this fit actually does not make much sense.
 It must be bounded by $2/3\leq \delta \leq 2$.
 The full red line
 would correspond to $c=20$ in \cite{Non} and $\delta(D)$. It is completely at odds with experiment. For this value of the parameter $c$ and experimental values of $E_R/U_0$, the model of \cite{Non} yields only super-ballistic Richardson diffusion.
 }
\label{Fig1}       
\end{figure}

Let us now clarify if the model in Refs. \cite{Non,Kessler10} can be supported
by experiments with cold atoms. One of such recent experiments \cite{Sagi12}
reveals sub-ballistic superdiffusion which is characterized by a L\'{e}vy distribution
of the particles positions $W_{\rm central}(x,t)$ obeying the scaling 
$W_{\rm central}(x,t)=t^{-1/\delta}{\cal L}_{\delta}(z=x/t^{1/\delta})$, where ${\cal L}_{\delta}(z)$ is
a L\'{e}vy distribution with index $0<\delta<2$. In \cite{Sagi12}, diffusion 
is characterized by the square of the width
of $W(x,t)$ at its half-maximum. It must be emphasized that such a L\'{e}vy distribution
in any experiment has necessarily cutoffs, i.e. it is tempered or truncated. This can be clearly
seen e.g. in Figs. 1 and 5 of \cite{Sagi12}, where experimental distributions do not extend beyond several millimeters
from their center. Also the model in \cite{Non} clearly supports such cutoffs. For any
 properly tempered 
 L\'{e}vy distribution,  and even  for any other distribution ${\cal L}_{\delta}(z)$ with finite
 second moment, $\langle \delta x^2(t)\rangle$ is finite and  proportional to 
 its squared width, at any time. This is just due to the experimentally observed scaling. Therefore,
 it would be reasonable to conclude that also experimentally $\langle \delta x^2(t)\rangle \propto t^{2/\delta}$
 and we can identify $\delta=2/\alpha(D)$ to compare the theory in \cite{Non} and the
 experiment in \cite{Sagi12}. This comparison is shown in Fig. \ref{Fig1}, with 
 $\delta(D)=4D/(7D-1)$, where $D=c E_R/U_0$,  $U_0$ is the optical lattice depth and
 $E_R$ is the recoil energy, with 
 $c$ being a single fitting parameter. Even the best fit with $c\approx 3.74$  is not
 acceptable, not saying already about $c\approx 20$ suggested in \cite{Non}. 
 However, in \cite{Non} the central part of $W(x,t)$ is given by the L\'{e}vy distribution
 with another index 
 \begin{eqnarray}
 \nu(D)=\frac{1}{3}+\frac{1}{3D}
 \end{eqnarray}
 instead of our $\delta(D)$. Notice that only for two values of $D$, $D=1/5$ (normal diffusion)
 and $D=1$ (Richardson diffusion), $\nu=\delta$, which has dramatic consequences, see below.
 Also fitting the experimental data with $\nu(D)$ instead of $\delta(D)$ does not help, see
 in Fig. 1. Even for the optimal value $c\approx 3.48$ in the corresponding fit, the theory does not match experiment.

 Furthermore, an interesting aspect of the theory in \cite{Non} is that the tail of $W(x,t)$, which
 is named infinite density therein, has a very different scaling from the  central part of 
 $W(x,t)$. This tail is scaled as 
 $W_{\rm tail}(x,t)=t^{-1-1/(2D)}{\cal I}(z=x/t^{3/2})$, where ${\cal I}(z)$ is a scaling function obtained in Ref. \cite{Non}. Namely this scaling yields Eq. (\ref{diff}).
 However, the experimental data in \cite{Sagi12}
 do not seem to support such a tail. The found in experiment scaling is very different.
 Most strikingly, the theory in \cite{Non} implies  that the diffusional spread of $W(x,t)$
 defined by its squared width at the half-maximum should be very different from the diffusional
 spread of the variance of the particles position.  If found experimentally, this very
 unusual behavior would mean that the very definition of  anomalous diffusion would heavily depend on   how to define the width of $W(x,t)$. For example, for $D=1/3$, which corresponds to the 
 ballistic diffusion in a standard definition with spatial variance, $\nu=4/3$, and $2/\nu=3/2$, which would correspond
 to sub-ballistic diffusion  from another point of view. For $D=0.4$
 in Fig. 1 of \cite{Non}, $\nu=7/6$, which still corresponds to sub-ballistic diffusion from the
 alternative point of view. However, in this case particles heat up to infinity and diffusion
 is clearly superballistic from the standard point of view of the spreading spatial variance.
 %Normal diffusion in 
 %from the point of view of $\langle \delta x^2(t)\rangle$ may be looking anomalous from the point
 %of view of the diffusional spread of the width of $W(x,t)$ at its half-maximum, 
 %and \textit{vice versa}. 
 I do not think, however, that such a strikingly unusual
  ambiguity of interpretation has ever been found experimentally.

 %Notice that this scaling is mono-fractal only for $D=1$,
 %which corresponds to Richardson diffusion with linearly growing in time kinetic
 %temperature $T_{\rm kin}(t)=\langle v^2(t)\rangle \propto t$, as it follows from
 %the results in \cite{Kessler10}.

To conclude, the applicability of the model in \cite{Non,Kessler10} to the systems of 
cold atoms is questionable not only in the heating superballistic phase $D\geq 1/3$, but also
overall.

Funding by the Deutsche Forschungsgemeinschaft, Grant
GO 2052/3-1 is gratefully acknowledged.

\end{document}